\documentclass{iopart}
\usepackage{graphicx}

\begin{document}

\newcommand{\be}{\begin{equation}}
\newcommand{\ee}{\end{equation}}

\newcommand{\bea}{\begin{eqnarray}}
\newcommand{\eea}{\end{eqnarray}}

\newcommand{\azAngle}{\varphi}
\newcommand{\EctSub}{{\rm CT}}


\article{QFEXT '05 - Quantum field theory under the influence of
  external conditions}{Quantum stabilization of Z-strings in the
  electroweak model}  
\author{O.~Schr\"oder} 
\address{School of Mathematics and
  Statistics, University of Plymouth, Plymouth PL4 8AA, UK}
\ead{oliver.schroder@plymouth.ac.uk}
\date{\today}
\begin{abstract}
\noindent We study the quantum energy of the Z-string in 2+1
dimensions using the phase shift formalism. Our main interest is
the question of stability of a Z-string carrying a finite fermion
number. 
\end{abstract}

\section{Introduction and motivation\label{sec_intro}}
Z-strings were first discovered as solutions of the classical field
equations of the electroweak model by Nambu \cite{Nambu} in the context
of bound pairs of magnetic monopoles. Later on they were rediscovered
- as independent objects in their own right - by Vachaspati
\cite{Vachaspati:1992fi}. 
\\
The main point under investigation in our study of Z-strings is their
stability. If they are stable, they would be relevant for a variety of
reasons: first of all, they would be the first solitonic objects in
the Standard Model to be found; given the importance and ubiquitousness of
solitonic objects in effective field theories in general it is
surprising that they seem to play no role in the Standard Model.
A second observation that might make them relevant is an alternative
scenario of (electroweak) baryogenesis proposed by Brandenberger et al
\cite{Brandenberger}. The presence of networks made from Z-strings
would make the requirement (for baryogenesis to happen at the
electroweak transition) of a first-order electroweak phase
transition obsolete. This is an attractive scenario since the
electroweak transition is known not to be of first order.
A third reason for
studying their stability is that - since Z-strings end in magnetic
monopoles - they might contribute to the primordial magnetic field. 
For a general overview of applications and properties of Z-strings
along with a large collection of references
cf. \cite{VachaspatiReview}. 

The structure of this paper is as follows: in section
\ref{sec_stability} we briefly discuss the notion of stability in the
presence of a conserved quantum number. Then, in section \ref{sec_tp},
we discuss the model under consideration, our method for computing the
fermion determinant and some thoughts necessary to choose parameters
properly in the D=2+1 dimensional theory. In section
\ref{sec_zstrings} we present the gauge and Higgs field ans\"atze used
for computing the fermion determinant. Then in section
\ref{sec_results} we present the (preliminary) results available at
the time of the QFEXT'05 conference. In section \ref{sec_conclusions}
we present an outlook and some (preliminary) conclusions.

\section{Stability\label{sec_stability}}
The question of stability is of prime importance in gauging the
importance of Z-strings. At this point one should keep in mind that
the Standard Model does not provide a topological stabilization
mechanism for string-like objects, unlike e.g. the Abelian Higgs
model. Hence every extended object has to be stable on energetic grounds.
In the purely bosonic, classical sector of
the electroweak model, Z-strings are solutions of the classical equations of
motion. But these solutions represent only a saddle point of the classical
energy functional and not a minimum. Hence, the Z-strings can decay,
e.g. by condensation of $\phi_+$ or $W$ bosons along the string - so
long as the value of the weak angle is close enough to its physical
value; in the unphysical region $\sin \Theta_w > 0.9$ the Z-string
actually is - classically - stable. These issues are discussed in 
depth and detail in the excellent review \cite{VachaspatiReview}.
If one considers fermions in additions to the bosonic sector of the
electroweak model, one finds that the Z-string actually binds fermions
to its core, some of them rather tightly \cite{VachaspatiReview}. 
Hence, one can investigate a new kind of stability
\cite{Khemani:2003}:  one can compare the total energy of the Z-string 
plus $N$ bound fermions to the energy of $N$ free fermions. If it is 
less, one has certainly found an interesting object, since even if the
configuration under investigation decays further, it cannot simply
decay to the vacuum (as it might in the absence of occupied bound
states) since the fermion number is conserved and it has already been
established that a configuration exists with energy below $N$ times the free
fermion number.
However, if one wants to consider the bound state energy of fermions
one also has to take into account the fermion determinant, since it
arises at the same order of an $\hbar$ (loop) expansion as the fermion
bound state energy.
Thus the quantity which we consider in this talk is the difference
$\Delta E(N)$ between the Z-string energy including the energy of $N$
bound fermions and the energy of $N$ free fermions:
\be
\Delta E(N) = E_{class}(\textrm{Z-string}) + E_{vac}(\textrm{Z-string}) +
\sum_{j=1}^N(|\omega^{b.s.}_j| - m). \label{eqDeltaEN}
\ee 
Here $E_{class}$ denotes the classical (bosonic) energy, $E_{vac}$
denotes the renormalized vacuum polarization energy originating from 
the fermion determinant including effects of the counterterms and   
$\omega_j^{b.s.}$ denote the bound state energies. The sum runs over
the occupied levels only. The fermion number of the Z-string is $N$.
 
\section{Technical prerequisites\label{sec_tp}}

The model we consider for our computations differs in some respects
from the full electroweak model. \textit{First} of all, we consider
our fermionic weak
iso-doublets to be degenerate in mass. This is the most serious of our
simplifications and cannot easily be gotten rid of, since without the
isospin symmetry the problem doesn't have enough symmetry to allow a
partial wave decomposition which is at the heart of our approach to
computing the fermion determinant\footnote{This channel decomposition
  also applies to the bound state part of the spectrum. In each
  channel we have a finite number of bound states which can be determined
using e.g. a shooting method.}. As a \textit{second}
simplification, we drop the hypercharge field from our consideration
and only consider the SU(2) gauge field. 
This seems to be an innocuous simplification and we have techniques to
deal with the U(1) field. A \textit{third} simplification that is used only to
simplify the algebra is the restriction of our fermionic sector to one
weak iso-doublet. The \textit{fourth} and final simplification is that we
consider the model in D=2+1 dimensions\footnote{Nonetheless we use
  Dirac four-spinors to describe the fermions in our theory.}. 
Obviously, calculations in
D=2+1 are much simpler than in D=3+1 due to the simplified UV divergence
structure. And since this is only an exploratory investigation - to
see whether it worthwhile to do the vastly more complicated
calculation in D=3+1 - it seems sensible to keep things simple where
possible.
But if there was no connection between results in D=2+1 and
D=3+1, this investigation could not fulfil its purpose. Fortunately
there is some evidence that the behaviour of energies and energy
densities of string-like objects in D=2+1 can be a good guide to the
behaviour of energies (per unit length) and energy densities in D=3+1.
In our investigation of
electromagnetic flux tubes \cite{Graham:2004jb, Weigel:2005} we have 
learned that - given that the same 
renormalization conditions are used in D=2+1 and D=3+1 - the
renormalized quantum energies are indeed very similar in their functional
dependence on widths and fluxes. We use as a working hypothesis that
the same is true in the electroweak model.  

The techniques used for expressing the vacuum polarization energy
(which is related to the fermion
determinant by dividing out the time interval $T$ for which the
determinant is evaluated) in terms of phase shifts from an
associated scattering problem have been described extensively in
\cite{Leipzig} and shall be outlined here only briefly. The vacuum
polarization energy can be renormalized effectively by
realizing that if one replaces the full phase shifts by their Born
approximation (of n$^{th}$ order) one gets the same result as by
restricting the full one-loop vacuum polarization energy to the sum of
Feynman diagrams with n external insertions of the background fields.
Hence, in D=2+1, the vacuum polarization energy is given by  
\bea \fl  E_{\rm vac} = - \frac{1}{2} \sum_{b.s.} (|\omega^{b.s.}_j|-m)
- \frac{1}{2}\int dk (\sqrt{k^2+m^2}-m) \sum_{M} \frac{1}{\pi}\,
  \frac{d}{dk}\Big[\delta_{M} - \sum_{n=1}^N \delta_{M}^{(n)}
  \Big] \nonumber \\  +  \sum_{n=1}^N E_{FD}^{(n)} +
  E_\EctSub, \label{eqVacPolEn}\eea
where $\omega_j^{b.s.}$ denotes the fermion bound state
energies\footnote{The bound state energies are determined from the
  first-order form of the Dirac equation.}, $\delta_M$
the (full) phase shift in angular momentum channel $M$,
$\delta_{M}^{(n)}$ its n$^{th}$ Born approximation; $E_{FD}^{(n)}$ is the
  energy contribution computed from Feynman diagrams with n external
  legs and $ E_\EctSub$ is the energy resulting from the counterterms
  \footnote{The calculation of the vacuum polarization energy per
  unit length in D=3+1 uses the same phase shifts, Born
  approximations and bound state energies but different kinematical 
 factors \cite{GrahamInterface}. Of course also Feynman diagram and
 counterterm contributions are different.}. Note that both the $k$
integral on the one hand and the sum 
  of Feynman diagrams plus counterterms on the other hand are
  separately finite. This in particular distinguishes our
  investigation from the first computation of the fermion
  determinant in the background of a Z-string performed by Groves et
  al \cite{Groves:1999ks} where 
  determinant and counterterms were individually divergent functions
  of the (proper-time) cut-off parameter and a finite result was only
  obtained by combining these two quantities. Alas, they are known
  only numerically, hence this procedure is numerically not
  stable. The other difference is that we focus on occupying bound
  states, whereas Groves et al were mainly concerned with computing
  the fermion determinant.

A last topic that merits discussion here is the question on how to choose
the parameters of our model. In 3+1 dimensions, the gauge, Yukawa,
Higgs self coupling and vacuum expectation value of the Higgs field
can straightforwardly expressed using the fermion mass, the Higgs
mass, the tree level mass of the W boson (the W boson mass at one-loop level
is a prediction based on the tree level mass because of our choice of
renormalization conditions) and the Fermi coupling $G_F$. In 2+1
dimensions, the masses can be unambiguously re-used; however, it is not
entirely obvious how to choose the Fermi coupling. It is ultimately
related to a cross section - a concept that would need some translation
into two spatial dimensions.

The aim of our 2+1 dimensional calculation is to be a guide to a
full 3+1 dimensional calculation. For the vacuum polarization energy
we have assured this by using the same renormalization conditions as
we'd use in a D=3+1 calculation.

We now fix the parameters in such a way that this is also true for the
classical energy: in D=2+1 we compute an energy $E^{2+1}_{class}$, but
in D=3+1 we compute an energy per unit length,
$E^{3+1}_{class}/L$. Hence, we require 
\be
E^{2+1}_{class} = \frac{E^{3+1}_{class}}{L} \times \textrm{fundamental
\ length},
\ee
where the fundamental length is given by half the Compton wave
length of the fermion we integrate out. This makes sense from a
physical perspective, since the fermion integrated out sets the scale
beyond which spatial structures cannot be resolved any more. Hence a
volume of thickness half the Compton wave length of this fermion is
`perceived' as a surface. Also in the QED flux tube computations a
relative factor of $\pi/m$ appeared between the D=2+1 dimensional
energies and D=3+1 dimensional energies per unit length.
This prescription now allows to express the model parameters in terms
of the physical parameters of the theory in D = 3+1. The question on
how to choose an appropriate $G_F$ in D=2+1 can thus be avoided.

\section{Z-strings\label{sec_zstrings}}

We compute the vacuum polarization energy for Higgs and gauge field
configurations of a specific form\footnote{In absence of the
  hypercharge field the Z-field reduces to the
  $W^3$-field. $\vec{W}^+$ is the conventional charged $W$ field,
  $\phi_0$ denotes the neutral Higgs field and $\phi_+$ the
  charged Higgs field.}:
    \bea 
    \phi & = \biggl( \begin{array}{c} \phi_+\\ \phi_0   
    \end{array} \biggr) = v \biggl( \begin{array}{c} -i
      f_H(\rho) \cos \xi_1 + f_P(\rho)\\   
      f_H(\rho) \sin \xi_1 e^{i \azAngle}
    \end{array} \biggr), \nonumber \\
    g \vec{W}^3 & =   \frac{\hat{\azAngle}}{\rho} 2 
    f_G(\rho)\sin^2 \xi_1, \nonumber \\ 
    \frac{g}{\sqrt{2}} \vec{W}^+ & = \frac{i \hat{\azAngle}}{2 \rho} e^{-i 
      \azAngle}  f_G( \rho ) \sin 2 \xi_1. \label{eqSphalSq}
    \eea
Here $\rho$ denotes the two-dimensional radius, $\rho =
\sqrt{x^2+y^2}$, and $\hat{\azAngle}$ is the unit vector in azimuthal direction.
The Higgs vacuum expectation value is given by $v$. The gauge coupling
is given by $g$. The functions $f_H(\rho), f_P(\rho)$ and $f_G(\rho)$ are
  \textit{profile functions} and are discussed in the remainder of
  this section. 
The requirement of finite classical energy necessitates for $\rho \to
0$ both $f_H \to 0$ and $f_G \to 0$. For $\rho \to \infty$, $f_G \to 1$ and
$\phi^{\dagger} \phi = |f_H|^2 + |f_P|^2 \to 1$ are required. 
As $\xi_1$ is changed from $\frac{\pi}{2}$ to $0$ the configuration is
changed continuously from the Z-string to a purely scalar
configuration without winding. Note that in this process the gauge
invariant length $\phi^{\dagger} \phi$ of the Higgs field does not
change since it is independent of $\xi_1$.
The classical energy is a continuous
function of $\xi_1$ which - for $f_P \equiv 0$ - has a maximum at
$\xi_1 = \pi/2$ and decreases continuously as $\xi_1 \to 0.$ This
illustrates our earlier statement that the Z-string is classically
unstable and can unwind without hitting a topological barrier.

The ansatz shown in \eref{eqSphalSq} is a subset of a more general ansatz
called the \textit{sphaleron square}, cf. also \cite{Klinkhamer:1994uy}.

 For our numerical investigations we cannot deal
with general profile functions, but rather need also ans\"atze for the
functions that have the proper behaviour for small and large $\rho$:
\bea
f_H(\rho) = 1 - e^{-\frac{\rho}{w_H}}, &\  f_P(\rho) = a_P e^{
-\frac{\rho}{w_P}}, &\  f_G(\rho) = 1 -
e^{-\frac{\rho^2}{w_G^2}}. 
\eea
Altogether we have four parameters (plus $\xi_1$): three widths $w_H,
w_G, w_P$ and one amplitude $a_P$ - the amplitudes for $f_H, f_H$ are
fixed by requirements of finite classical energy mentioned above.

\section{Results\label{sec_results}}

In this section we want to present a couple of preliminary results for
classical, vacuum polarization and bound state energies. The plots
have in common that we have fixed $\xi_1 = \frac{\pi}{2}$, i.e. we present
results for the Z-string configuration. 

Couplings and Higgs vacuum expectation value are determined by our
choice of masses and the Fermi coupling. We choose for the fermion
mass 170 GeV, for the Higgs mass 115 GeV, for the tree level W boson
mass 80 GeV and for the Fermi coupling $10^{-5} \mathrm{GeV}^{-2}$.

When studying the plots the following point has to be kept in mind:
since $\xi_1$ is fixed, we have a four-parameter numerical
problem. The plots are only standard two-dimensional plots, hence
energies can only be plotted as a function of a single variable. We
have chosen NOT to fix the other parameter values but plot the
energies of all the 
configurations that we have available. Hence for each observable there
is not a curve but a band corresponding to the ranges of the variables
not represented on the x axis of the plot.
\begin{figure}
\centerline{
\includegraphics[width=7cm, height = 8cm, angle = 270]{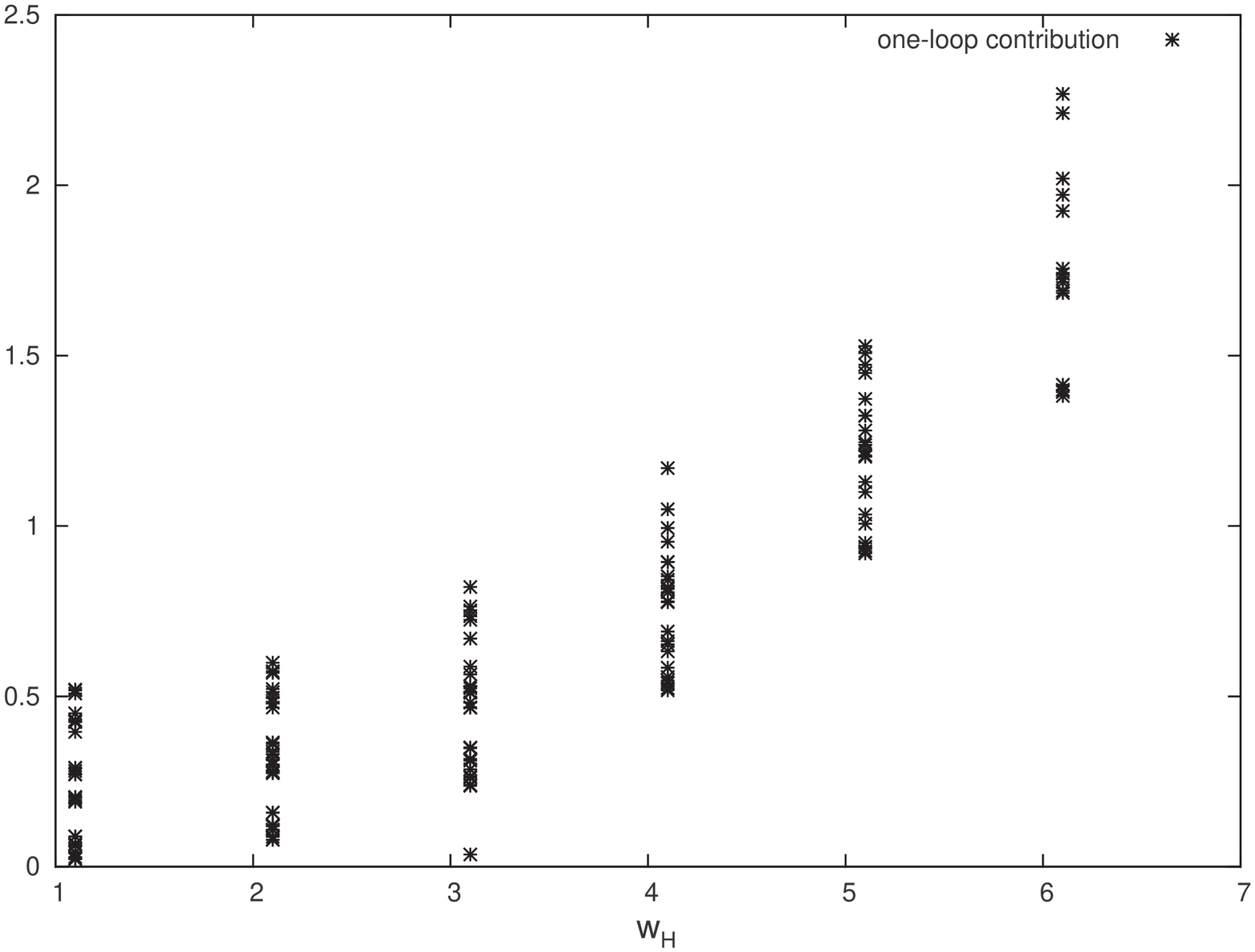} \hskip0cm
\includegraphics[width=7cm, height = 8cm, angle = 270]{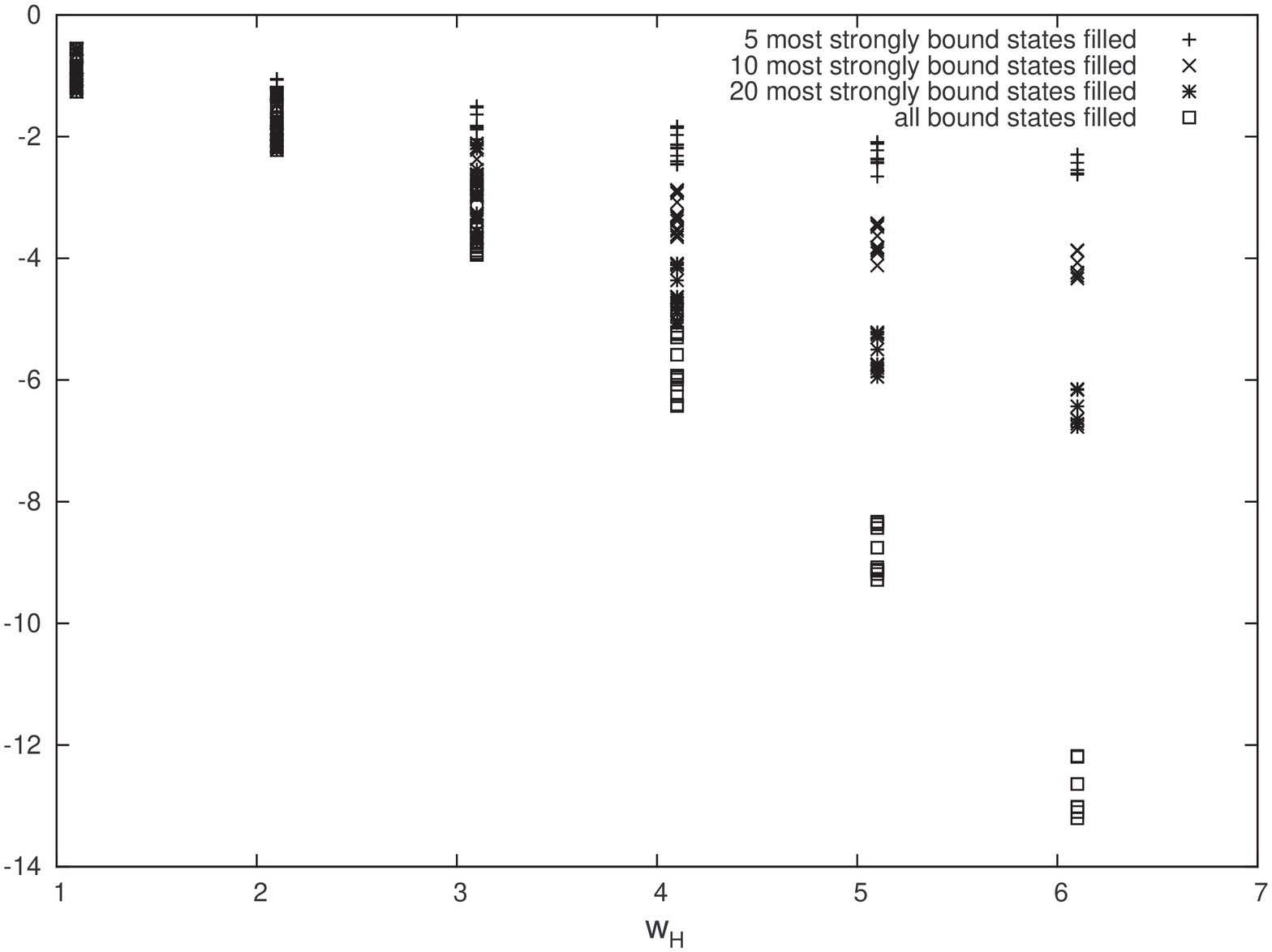}}
\caption{\label{figure1} \sf The 
left panel shows the (completely renormalized) vacuum
  polarization energy as a function of the width $w_H$, the right
  panel shows the energy gained by filling bound state levels relative
  to the same number of free fermions,
  $\sum_{j=1}^{N}(|\omega^{b.s.}_j|-m)$ for N=5,10,20 and all
  available bound states. Note that this latter curve
corresponds to comparing configurations with different fermion number.}
\end{figure}
In figure \ref{figure1} we show in the left panel the renormalized
vacuum polarization energy (in units of the fermion mass) including
Feynman diagrams and counterterm 
contributions as function of the width of the neutral Higgs field,
$w_H$. This seems to be the predominant dependence. The contents of
the right panel are slightly more difficult to explain:
in \eref{eqDeltaEN}  only one part of $\Delta E(N)$ depends on the
fermion number of the configuration, namely
\be
\Delta E^{b.s.}(N) = \sum_{j=1}^{N} (|\omega^{b.s.}_j| - m). \label{eqBSgain}
\ee
Since we look for a stable object we restrict our choice of bound
states in \eref{eqBSgain} to the most strongly bound states. In the
right panel of figure \ref{figure1} we now plot $\Delta E^{b.s.}(N)$ for N=
5, 10 and 20. For the lowest curve we
occupy \textit{all available} bound states, denoted in the following
as $\Delta E^{b.s.}(\mathrm{all})$. Thus points along this
curve can and will correspond to different fermion numbers. Whereas the
curves with fixed 
fermion number level off as $w_H$ increases this latter curve keeps
decreasing as $w_H$ increases. This is due to an amazing proliferation
of bound states. Whereas for $w_H \approx 2$ there is only a handful
of bound states, for $w_H \approx 6$ there can be easily more than 80
bound states. The dependence of the number of bound states seems to be
roughly quadratic - as $w_H$ increases, both more and more angular momentum
channels contain bound states and the number in each individual channel keeps
increasing, too.
More recent investigations, performed after QFEXT'05,
show that this behaviour continues at least up to widths $w_H \approx
12$. In this parameter regime we have found configurations with around
500 bound states in various angular momentum channels.

\begin{figure}
\centerline{
\includegraphics[width=8cm, height =14cm, angle = 270]{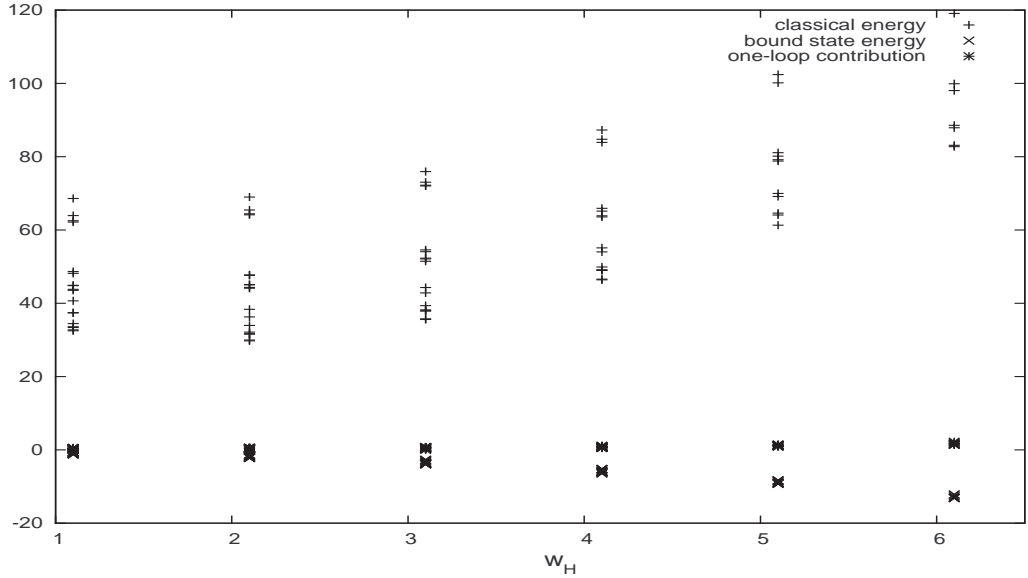} 
}
\caption{\label{figure2}\sf The figure shows the classical
  energies (upper band of points), vacuum polarization energies
  (denoted 'one-loop 
  contributions', medium band of points) and $\Delta
  E^{b.s.}(\mathrm{all})$ (lower band of points) as a function of the
width $w_H$. }
\end{figure}

In figure \ref{figure2} we plot the classical energy, the vacuum
polarization energy and $\Delta E^{b.s.}(\mathrm{all})$. This figure
shows clearly the different orders of magnitude that are
involved. Furthermore it can be seen clearly that the effect of
populating bound states by far outweighs the increase in energy due to
taking into account the fermion determinant.

When combining the classical energy with the vacuum polarization energy and
$\Delta E^{b.s.}(N)$ to form $\Delta E(N)$, one has to keep in mind
that the fermions - in contrast to the bosons in this model - carry a
colour quantum number and that fermions with different colour are
energetically degenerate. Hence, both $E_{vac}$ and $\Delta
E^{b.s.}(N)$ have to be multiplied by the number of colours $N_C$
before they are added to $E_{class}$. The fermion number under
consideration then is actually $N \times N_C$.

\begin{figure}
\centerline{ \includegraphics[width=6cm, height =
  5cm, angle = 270]{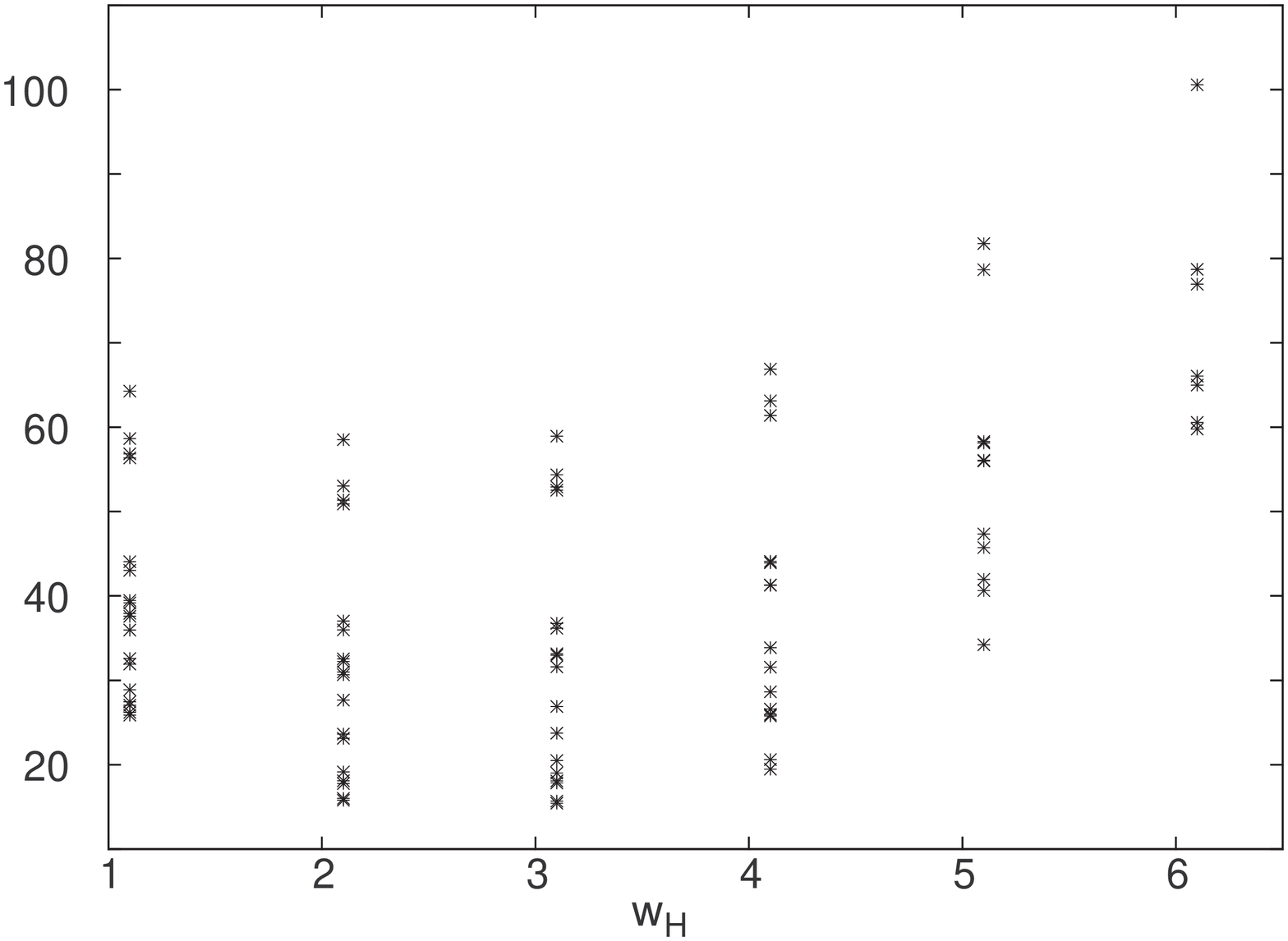} \hskip0.5cm
\includegraphics[width=6cm, height =
5cm, angle = 270]{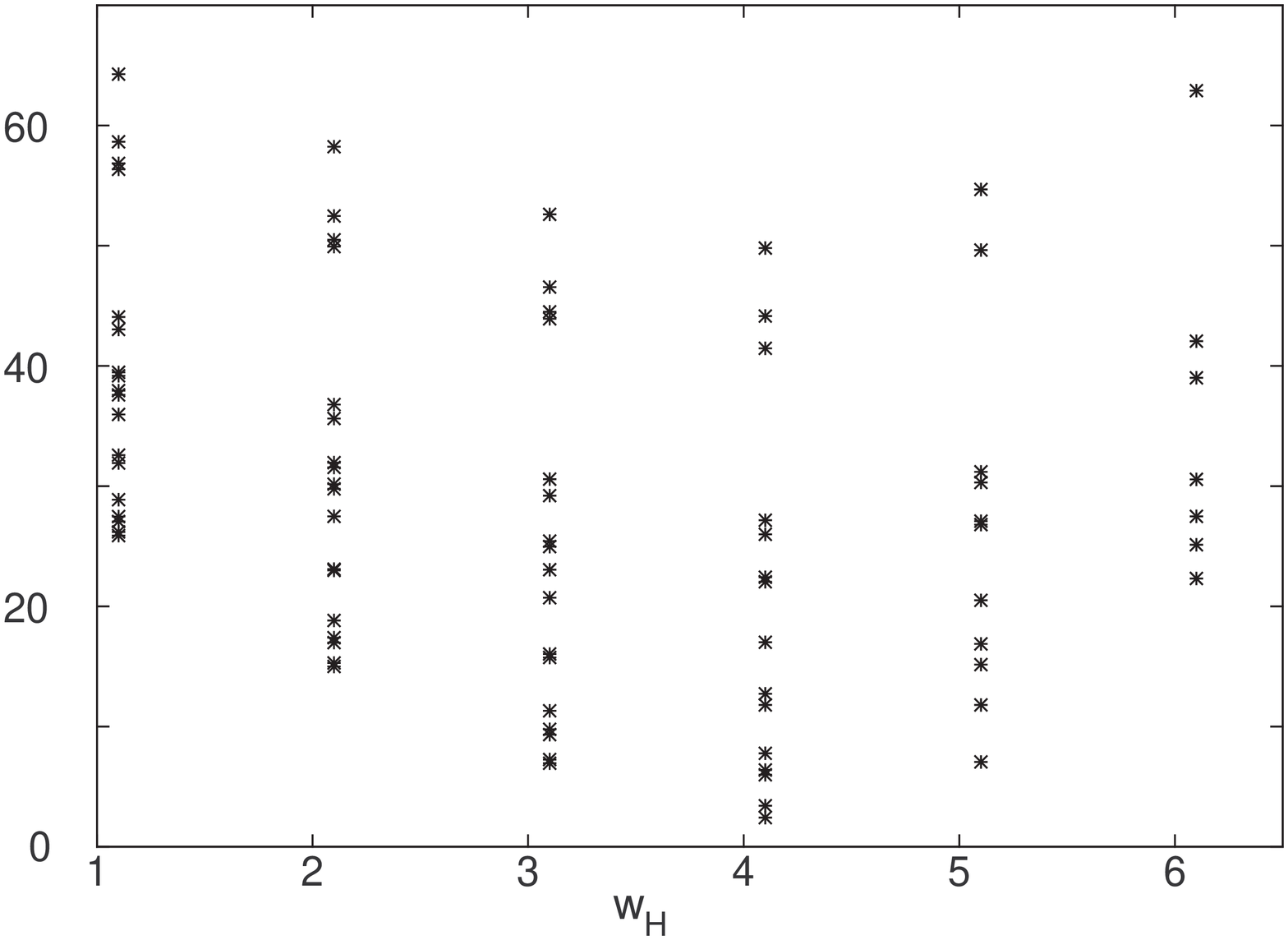} \hskip0.5cm 
\includegraphics[width=6cm, height = 5cm, angle = 270]{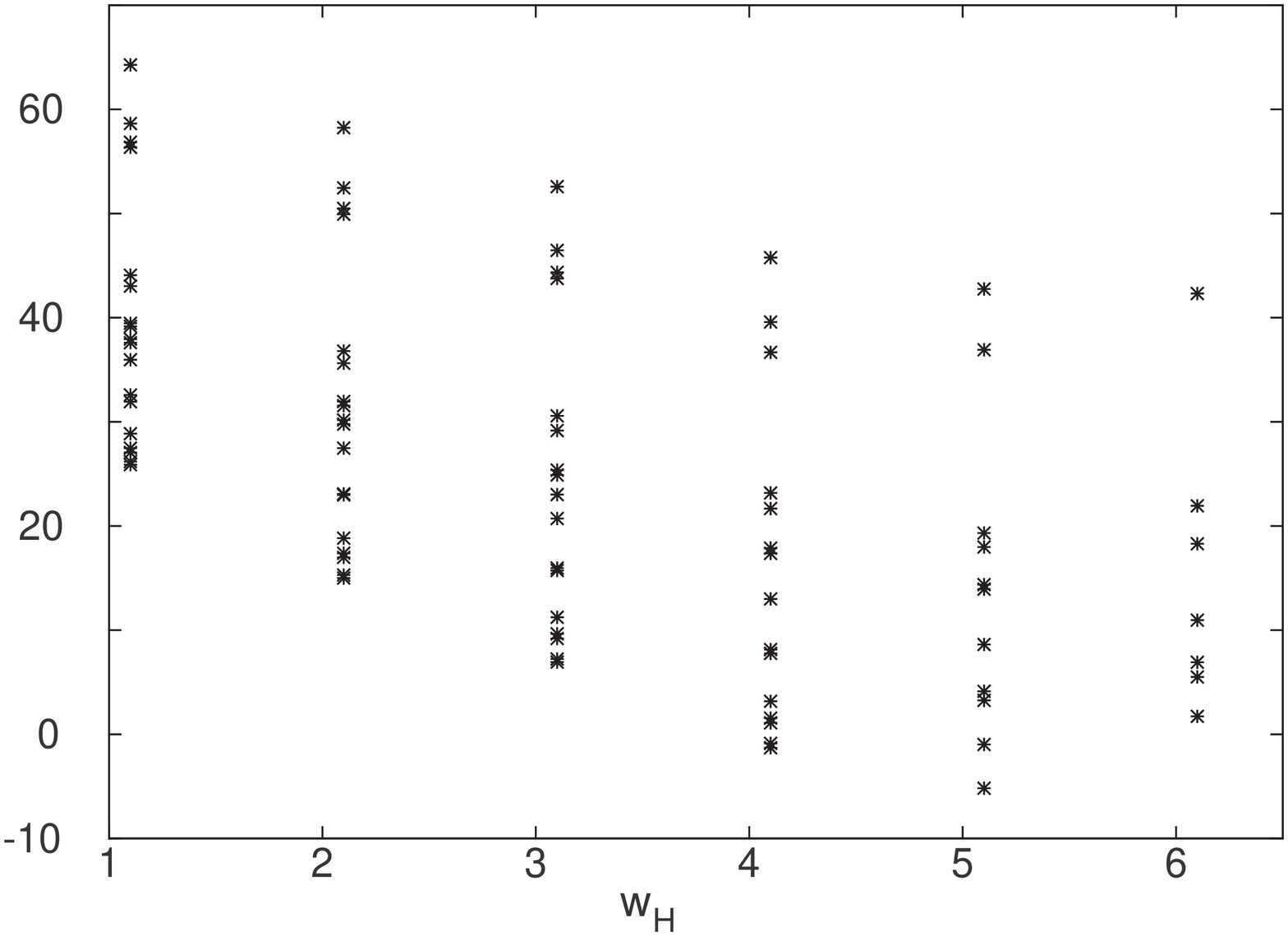}}
\caption{\label{figure3}\sf  In this figure we present our results for
  $\Delta E(N)$ as a function of the width $w_H$ for different values of $N$,
  from left to right $N=10, 30, 50$. The actual fermion numbers are
  (again from left to right) $90, 270, 450$.}
\end{figure}

In figure \ref{figure3} we use $N_C=9$. In the left panel of figure
\ref{figure3} we have
filled the 10 most strongly bound states. We find a minimum, but since
we plot $\Delta E$ this minimum has to be below zero to indicate a
stable object. So for both $N = 10$ and $N = 30$ (panel in the
middle) the object under consideration is not stable. The situation is
different for $N = 50$, since the minimum there is clearly below
zero. Since $N$ does not include the colour degeneracy this stable
object actually carries fermion number 450.

Also, a certain pattern seems to emerge from figure \ref{figure3}:
as $N$ increases, the minimum of $\Delta E(N)$ moves towards larger
values of $w_H$ and the value of $\Delta E(N)$ at the minimum
decreases. This is of course due to the fact that the larger $w_H$ the
more actual bound states are available and hence the possible gain in
energy by filling these bound states also increases. It may even be
possible that for sufficiently large $w_H$ the energy gain is large enough
to allow a stable object even for $N_C = 3$ to exist, but this is a
question under current investigation and cannot be answered at the
moment.

A different investigation - results will have to be reported elsewhere
- considers a heavy fermion with masses around $1.5 TeV$ instead of
the top quark mass used here. 

\section{Conclusions\label{sec_conclusions}}
For this contribution only preliminary data were
available. Nevertheless we can state that - given a sufficient number
of colours - we have found a very interesting object that is stable in
the sense of section \ref{sec_stability}. It is very large and carries a
gigantic fermion number 
(450). At the moment it is not clear what happens if we increase the
size of the object further - will we find so many bound states that
maybe it is possible to find a stable object even for $N_C = 3$? 
Therefore a in-depth investigation of the parameter space is urgently
needed, and is already under way \cite{Graham:2006}. Also, the great
effort to investigate the D=3+1 case is now fully justified.

\ack
First of all I'd like to thank my collaborators N. Graham, V. Khemani,
M. Quandt and H. Weigel for all the effort that they've put into the
Z-string project (among other things). Furthermore, I'd like to thank
the organizers of QFEXT'05 for the marvellous job they have done in
making the conference happen and the other participants for many
interesting conversations. Also, I'd like to thank the Plymouth
Particle Theory Group for stimulating discussions.
Some parts of the calculation have been
simplified tremendously by FORM \cite{Vermaseren:2000nd} and
\textsc{Mathematica}\cite{Mathematica}.   
This project has been supported in part by the \textit{Deutsche
  Forschungsgemeinschaft} under contract Schr 749/1-1 and the
\textit{Particle Physics and Astronomy Research Council}.

\end{document}